\documentclass[prl,twocolumn,showpacs,preprintnumbers,amsmath,amssymb,superscriptaddress]{revtex4-1}

\usepackage{graphicx}
\usepackage{dcolumn}
\usepackage{bm}
\usepackage{amsthm}
\usepackage{amsmath}
\usepackage{amssymb}
\usepackage{xcolor}

\begin{document}

\title{Resolving complex spin textures in nanoparticles by magnetic neutron scattering}

\author{Laura~G.~Vivas}\email[Electronic address: ]{lauragvivas@gmail.com}
\affiliation{Department of Physics and Materials Science, University of Luxembourg, 162A~avenue de la Fa\"iencerie, L-1511 Luxembourg, Grand Duchy of Luxembourg}
\author{Rocio~Yanes}
\affiliation{Department of Applied Physics, University of Salamanca, Salamanca, 37008, Spain}
\author{Dmitry~Berkov}
\author{Sergey~Erokhin}
\affiliation{General Numerics Research Lab, Moritz-von-Rohr-Straße~1A, D-07745, Jena, Germany}
\author{Mathias~Bersweiler}
\author{Dirk~Honecker}
\author{Philipp~Bender}
\author{Andreas~Michels}\email[Electronic address: ]{andreas.michels@uni.lu}
\affiliation{Department of Physics and Materials Science, University of Luxembourg, 162A~avenue de la Fa\"iencerie, L-1511 Luxembourg, Grand Duchy of Luxembourg}

\date{\today}

\begin{abstract}
In the quest to image the three-dimensional magnetization structure we show that the technique of magnetic small-angle neutron scattering (SANS) is highly sensitive to the details of the internal spin structure of nanoparticles. By combining SANS with numerical micromagnetic computations we study the transition from single-domain to multi-domain behavior in nanoparticles and its implications for the ensuing magnetic SANS cross section. Above the critical single-domain size we find that the cross section and the related correlation function cannot be described anymore with the uniform particle model, resulting e.g.\ in deviations from the well-known Guinier law. We identify a clear signature for the occurrence of a vortex-like spin structure at remanence. The micromagnetic approach to magnetic SANS bears great potential for future investigations, since it provides fundamental insights into the mesoscale magnetization profile of nanoparticles.
\end{abstract}

\maketitle\

{\it Introduction.} A key challenge in magnetism remains the visualization of complex three-dimensional magnetization vector fields in the bulk of materials~\cite{Pacheco2017}. Recent progress in this direction has been made by \textcite{Donnelly2017,Donnelly2020}, who have developed the technique of X-ray vector nanotomography which allows one to image individual magnetic structures with a lateral resolution of about $50 \, \mathrm{nm}$. Here, we demonstrate that magnetic small-angle neutron scattering (SANS) can be employed to resolve complex inhomogeneous spin textures in the complimentary size regime ($\sim 1-100 \, \mathrm{nm}$). This is particularly relevant e.g.\ for the study of the internal spin structure of magnetic nanoparticle systems~\cite{rmp2019}. 

Numerous magnetic SANS studies on nanoparticle ensembles point towards a strong deviation from the homogeneously magnetized single-domain state (see, e.g., Refs.~\cite{disch2012,kryckaprl2014,ijiri2014,guenther2014,maurer2014,dennis2015,grutter2017,oberdick2018,krycka2019,benderapl2019,bersweiler2019,kryckaprm2020} and references therein). A fundamental problem is, however, that the magnetic SANS data analysis is still in its infancy, since it largely utilizes structural form-factor models for the cross section, adapted from nuclear SANS, which fail to account for the existing spin inhomogeneity inside magnetic nanoparticles. For the analysis of experimental magnetic SANS data, the spatial nanometer-scale variation of the orientation and magnitude of the magnetization vector field $\mathbf{M}(\mathbf{r})$ must be taken into account, as has been demonstrated for nano\-structured \textit{bulk} ferromagnets~\cite{michels2013,michels2014jmmm,mettus2015,erokhin2015,metmi2015,metmi2016,michelsPRB2016,michelsdmi2019,mistonov2019}.

In this paper we numerically solve Brown's static equations of micromagnetics~\cite{brown} to monitor the transition from the single-domain to the multi-domain state in microstructural-defect-free spherical Fe nanoparticles. The ensuing results for the magnetic SANS cross section and correlation function reveal marked differences as compared to the superspin model and provide guidance for the experimentalist to identify nonuniform spin structures inside nanoparticles. The combination of numerical micromagnetics and magnetic SANS is a promising approach for the resolution of three-dimensional magnetization structures. We refer to the Supplemental Material (SM)~\cite{laura2019sm} for details regarding the micromagnetic simulations, which include all the relevant magnetic interactions such as the magnetodipolar interaction, Zeeman energy, magnetocrystalline anisotropy, and the isotropic exchange interaction. In the SM we also relate our simulation results to experimental SANS data on $38$~nm-sized Manganese-Zinc-Ferrite nanoparticles~\cite{bersweiler2019}.

{\it Magnetic SANS cross section and pair-distance distribution function.} The quantity of interest is the elastic magnetic differential scattering cross section $d \Sigma_M / d \Omega$, which is recorded on a two-dimensional position-sensitive detector. For the most commonly used scattering geometry in magnetic SANS experiments, where the applied magnetic field $\mathbf{H}_0 \parallel \mathbf{e}_z$ is perpendicular to the wave vector $\mathbf{k}_0 \parallel \mathbf{e}_x$ of the incident neutrons (see Fig.~1 in \cite{laura2019sm}), $d \Sigma_M / d \Omega$ for unpolarized neutrons can be written as~\cite{rmp2019}:
\begin{eqnarray}
\frac{d \Sigma_M}{d \Omega} = \frac{8 \pi^3}{V} b_H^2 \left( |\widetilde{M}_x|^2 + |\widetilde{M}_y|^2 \cos^2\theta \right. \nonumber \\ \left. + |\widetilde{M}_z|^2 \sin^2\theta - (\widetilde{M}_y \widetilde{M}_z^{\ast} + \widetilde{M}_y^{\ast} \widetilde{M}_z) \sin\theta \cos\theta \right) ,
 \label{eq:equation1}
\end{eqnarray}
where $V$ is the scattering volume, $b_H = 2.91 \times 10^8 \, \mathrm{A}^{-1}\mathrm{m}^{-1}$ is the magnetic scattering length in the small-angle regime (the atomic magnetic form factor is approximated by $1$, since we are dealing with forward scattering), $\widetilde{\mathbf{M}}(\mathbf{q}) = \{ \widetilde{M}_x(\mathbf{q}), \widetilde{M}_y(\mathbf{q}), \widetilde{M}_z(\mathbf{q}) \}$ represents the Fourier transform of the magnetization vector field $\mathbf{M}(\mathbf{r}) = \{ M_x(\mathbf{r}), M_y(\mathbf{r}), M_z(\mathbf{r}) \}$, $\theta$ denotes the angle between $\mathbf{q}$ and $\mathbf{H}_0$, and the asterisk ``$*$'' marks the complex-conjugated quantity. As shown in~\cite{laura2019sm}, for a uniformly magnetized spherical particle of radius $R$ with its saturation direction parallel to $\mathbf{e}_z$, Eq.~(\ref{eq:equation1}) reduces to:
\begin{eqnarray}
\frac{d \Sigma_M}{d \Omega} = V_p (\Delta \rho)_{\mathrm{mag}}^2 \, 9 \left( \frac{j_1(qR)}{qR} \right)^2 \sin^2\theta ,
\label{homomagsans2}
\end{eqnarray}
where $V_p = \frac{4\pi}{3} R^3$, $(\Delta \rho)_{\mathrm{mag}}^2 = b_H^2 \left( \Delta M \right)^2$ is the magnetic scattering-length density contrast, and $j_1(z)$ is the first-order spherical Bessel function.

The pair-distance distribution function $p(r)$ can be computed from the azimuthally-averaged magnetic SANS cross section according to:
\begin{eqnarray}
\label{pvonreqintegral}
p(r) = r^2 \int\limits_0^{\infty} \frac{d \Sigma_M}{d \Omega}(q) \frac{\sin(qr)}{qr} q^2 dq ,
\end{eqnarray}
which corresponds to the distribution of real-space distances between volume elements inside the particle weighted by the excess scattering-length density distribution; see the reviews by Glatter~\cite{glatterchapter} and by Svergun and Koch~\cite{svergun03} for detailed discussions of the properties of $p(r)$ and for information on how to compute $p(r)$ by indirect Fourier transformation~\cite{bender2017}. Apart from constant prefactors, the $p(r)$ of the azimuthally-averaged single-particle SANS cross section [Eq.~(\ref{homomagsans2})], corresponding to a uniform sphere magnetization, equals (for $r \leq 2R$):
\begin{eqnarray}
\label{pvonreq}
p(r) = r^2 \left( 1 - \frac{3r}{4R} + \frac{r^3}{16 R^3} \right) .
\end{eqnarray}
As is demonstrated in the following, when the particle's spin structure is inhomogeneous, the $d \Sigma_M / d \Omega$ and the corresponding $p(r)$ differ significantly from the homogeneous particle case [Eqs.~(\ref{homomagsans2}) and (\ref{pvonreq})].

\begin{figure*}[tb!]
\centering
\resizebox{2.0\columnwidth}{!}{\includegraphics{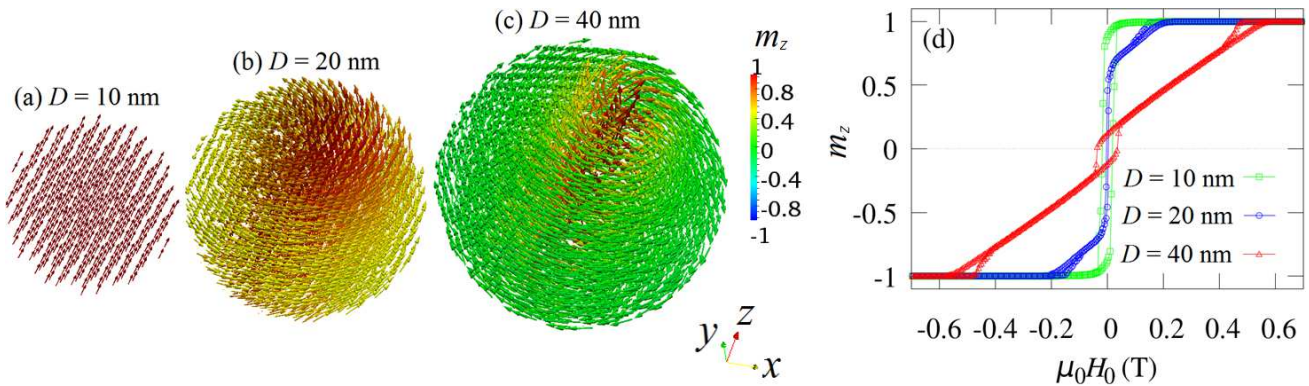}}
\caption{Numerically computed magnetization distributions of Fe nanospheres with diameters of (a)~$D = 10 \, \mathrm{nm}$, (b)~$D = 20 \, \mathrm{nm}$, and (c)~$D = 40 \, \mathrm{nm}$. For the micromagnetic simulations the software package MuMax3~\cite{mumax3new,leliaert2019} was used. Shown are the spin structures in the remanent state [(a) and (c)] and at the coercive field $H_c$~(b) after prior saturation along the $z$-direction. (d)~Corresponding normalized hysteresis loops (random particle orientations).}
\label{fig1}
\end{figure*}

{\it Results and Discussion.} Figure~\ref{fig1} displays the spin structures and magnetization curves of Fe spheres with diameters of $D = 10, 20$~and~$40 \, \mathrm{nm}$. As far as the typical resolution range of a SANS experiment is concerned ($\sim 1-100 \, \mathrm{nm}$), these three particle sizes cover the characteristic spin structures and magnetization-reversal mechanisms found on defect-free Fe spheres. Using the materials parameters specified in~\cite{laura2019sm} we find a critical single-domain diameter for Fe of $D_c \cong 72 \sqrt{A K_1} / (\mu_0 M_s^2) \cong 13.6 \, \mathrm{nm}$~\cite{mumagcoll,skomskibook,sergeycommentsd}. As one can see in Fig.~\ref{fig1}(a), the magnetization distribution of the $D = 10 \, \mathrm{nm}$ sphere is quasi-uniform in the remanent state, while the one of the $D = 40 \, \mathrm{nm}$ sphere is highly inhomogeneous, i.e., $\mathbf{M} = \mathbf{M}(x,y,z)$, and exhibits a vortex-type configuration [Fig.~\ref{fig1}(c)]. The $D = 20 \, \mathrm{nm}$ nanosphere is nearly homogeneously magnetized in the remanent state (data not shown), as the $D = 10 \, \mathrm{nm}$ sphere, but is highly inhomogeneous at the coercive field [Fig.~\ref{fig1}(b)]~\cite{aharonibook}. In the calculations we find (using the materials parameters of Fe) that nonuniform magnetization states appear at $H_0 = 0$ once $D$ is larger than roughly $20 \, \mathrm{nm}$.

The hysteresis loop of the $D = 10 \, \mathrm{nm}$ sphere very well reproduces the literature results for randomly-oriented Stoner-Wohlfarth particles with cubic anisotropy~\cite{usov1997}, resulting in a coercive field and reduced remanence of, respectively, $\mu_0 H_c = 0.33 \times 2K/M_s \cong 18 \, \mathrm{mT}$ and $M_r/M_s = 0.831$. On the other hand, the magnetization data for $D = 20 \, \mathrm{nm}$ sphere exhibit the nucleation and propagation of a domain, while the spin structure of the $40 \, \mathrm{nm}$ sample is characterized by the nucleation and propagation of a vortex. With respect to magnetic neutron scattering, the observations in Fig.~\ref{fig1} clearly suggest that for diluted scattering systems, corresponding to the single-sphere case, the macrospin model is not appropriate anymore beyond a certain particle size. In order to highlight this, we will now compare and discuss the $d \Sigma_M / d \Omega$ and $p(r)$ of the $D = 40 \, \mathrm{nm}$ particle with the one of a uniformly magnetized sphere of the \textit{same} size.

\begin{figure}[tb!]
\centering
\resizebox{1.0\columnwidth}{!}{\includegraphics{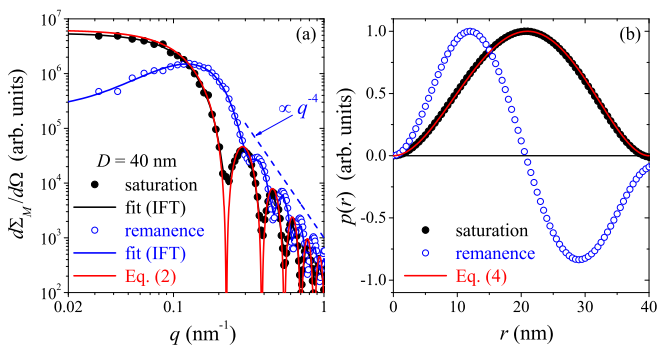}}
\caption{(a)~Comparison of the (over $2\pi$) azimuthally-averaged magnetic SANS cross sections $d \Sigma_M / d \Omega$ of a uniformly (closed black circles) and a nonuniformly (open blue circles) magnetized Fe sphere (remanent state) with a diameter of $D = 40 \, \mathrm{nm}$ (log-log scale). Both cross sections have been numerically computed using the continuum theory of micromagnetics. Solid black and blue lines:~Fit of $d \Sigma_M / d \Omega$ based on the indirect Fourier transform (IFT) from~(b). Solid red line:~analytical solution using the sphere form factor [azimuthally-averaged version of Eq.~(\ref{homomagsans2})]. Dashed blue line:~$d\Sigma_M / d \Omega \propto q^{-4}$. (b)~Corresponding distance distribution functions $p(r)$.}
\label{fig2}
\end{figure}

Figure~\ref{fig2}(a) depicts the azimuthally-averaged $d \Sigma_M / d \Omega$ of an assumed uniformly and a nonuniformly-magnetized sphere with a diameter of $D = 40 \, \mathrm{nm}$; see~\cite{laura2019sm} for the two-dimensional $d \Sigma_M / d \Omega$ and for the decryption of the cross section into its individual Fourier components. Based on the numerically computed $d \Sigma_M / d \Omega$ the corresponding correlation functions $p(r)$ [Fig.~\ref{fig2}(b)] were computed using the indirect Fourier transform (IFT) method~\cite{bender2017}. The $p(r)$ of the inhomogeneous sphere exhibits an oscillatory character with a zero crossing at $r \cong 20.7 \, \mathrm{nm}$, while the numerical $p(r)$ of the homogeneous sphere agrees excellently with the analytical result [Eq.~(\ref{pvonreq})], as expected. Compared to the homogeneous case, the oscillations in $d \Sigma_M / d \Omega$ for the inhomogeneous sphere are shifted to larger $q$-values, since the magnetization distribution is composed of structures smaller than $D$. If the $d \Sigma_M / d \Omega$ data in the remanent state would be fitted using the sphere form factor---assuming a uniform magnetization--- then an erroneously small value for the particle size may result. Moreover, since the inhomogeneous sphere is characterized by a reduced magnetization (remanence), the cross section in the limit $q \rightarrow 0$ is reduced as compared to the homogeneous case. This then implies that the famous Guinier law, which describes the low-$q$ region of a small-angle scattering curve ($d \Sigma_M / d \Omega \propto \exp(- q^2 R_G^2 / 3)$ with $R_G$ the radius of gyration), does not hold for an inhomogeneously magnetized particle (sphere). The Porod law, $d\Sigma_M / d \Omega \propto q^{-4}$ [dashed line in Fig.~\ref{fig2}(a)], is however even found for the inhomogeneous particle, since the spin structure is still confined by a sharp phase boundary. We emphasize that the Porod behavior naturally emerges from the micromagnetic computations without \textit{a priori} assumptions.

\begin{figure*}[tb!]
\centering
\resizebox{2.0\columnwidth}{!}{\includegraphics{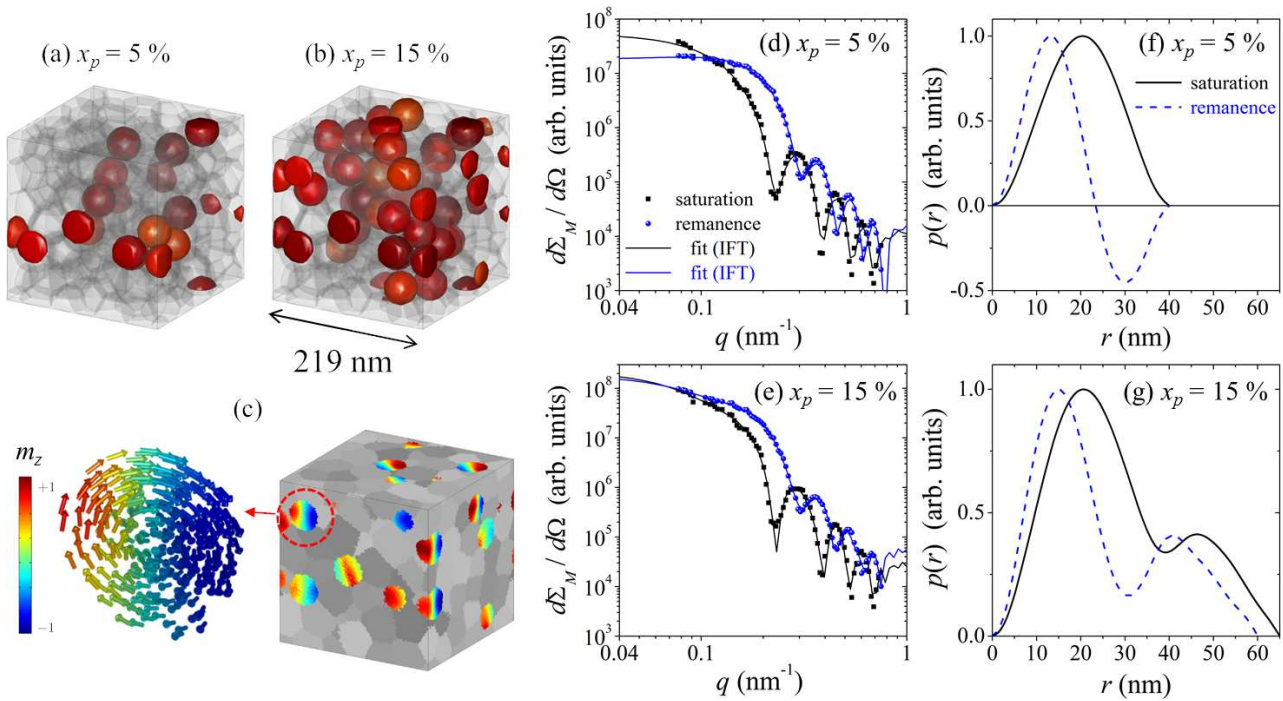}}
\caption{(a) and (b)~Microstructures used in the micromagnetic simulations of dense assemblies of $40 \, \mathrm{nm}$-sized Fe particles. (c)~Snapshot of the magnetic microstructure of the $x_p = 15 \, \%$ sample and of the spin distribution of a single Fe nanoparticle at remanence. (d) and (e)~Azimuthally-averaged $d \Sigma_M / d \Omega$ of Fe nanoparticles at saturation and at remanence and for different Fe nanoparticle volume fractions $x_p$ (see inset) (log-log scale). (f) and (g)~Corresponding distance distribution functions $p(r)$.}
\label{fig3}
\end{figure*}

Figure~\ref{fig3} shows the simulated sample microstructures and spin structures of dense assemblies of particles [Fig.~\ref{fig3}(a)$-$(c)] and the results for $d \Sigma_M / d \Omega$ [Fig.~\ref{fig3}(d) and (e)] and $p(r)$ [Fig.~\ref{fig3}(f) and (g)]. The simulations were performed on \textit{monomodal} $40 \, \mathrm{nm}$-sized spherical Fe particles of different volume fractions $x_p$ ranging between $5-15 \, \%$~\cite{laura2019sm,erokhinpra2017}. A particle size of $D = 40 \, \mathrm{nm}$ was chosen in order to compare the results on the interacting system (Fig.~\ref{fig3}) to the single-particle case (Fig.~\ref{fig2}), where for $D \sim 30-120 \, \mathrm{nm}$ a magnetic vortex structure is observed in the remanent state. This vortex-type spin configuration is also visible at the remanent state for all the concentrations used [e.g., Fig.~\ref{fig3}(c)]. Inspecting the magnetic scattering of the system with the lowest Fe particle concentration of $x_p = 5 \, \%$ [Fig.~\ref{fig3}(d)] we see, as with the single-particle cross section [Fig.~\ref{fig2}(a)], a marked difference between the $d \Sigma_M / d \Omega$ at saturation and at remanence; in particular, for $H_0 = 0$, the oscillations in $d \Sigma_M / d \Omega$ are shifted to larger $q$-values, which is due to the forming inhomogeneous spin texture. However, the low-$q$ behavior of $d \Sigma_M / d \Omega$ of the weakly interacting system is different than in the single-particle case:~while in the latter we have observed the absence of Guinier behavior ($d \Sigma_M / d \Omega$ decreases when $q \rightarrow 0$), the $d \Sigma_M / d \Omega$ for $x_p = 5 \, \%$ approaches a plateau, which we attribute to the emerging interparticle interactions. Increasing the particle concentration further to $x_p = 15 \, \%$ we still see the shift in the oscillations of $d \Sigma_M / d \Omega$ between saturation and remanence [Fig.~\ref{fig3}(e)], however, the magnetic SANS cross section at low $q$ now \textit{increases} with decreasing momentum transfer.

In liquid-state theory the behavior of the structure factor $S(q)$ in the limit of low $q$, when the system is probed on macroscopic length scales, provides information on thermodynamical quantities and on the nature of the interaction potential (repulsive vs.\ attractive) between the particles. Using the compressibility relation one can show that repulsive interparticle interactions result in $S(q = 0) < 1$, whereas attractive interactions give $S(q = 0) > 1$~\cite{chentartagliabook}. Translated to our magnetic problem, the observed change in the slope of $d \Sigma_M / d \Omega$ can be related to the increased strength of the dipolar interaction between the particles~\cite{honecker2020}. With increasing volume fraction $x_p$ the large-scale spin arrangement of the ensemble is, on the average, becoming more homogeneous due to the constraint of $\nabla \cdot \mathbf{M} = 0$ imposed by the pole-avoidance principle. Besides, as is seen in Fig.~6 in~\cite{laura2019sm}, the internal nanoparticle spin structure, which is probed on a larger $q$-range, is also becoming more homogeneous with increasing $x_p$. 

The inhomogeneous vortex-type spin structure at remanence is related to the first peak in $p(r)$, which shifts only very slightly from the single-particle case ($r^{\mathrm{rem}}_{\mathrm{max}} \cong 12 \, \mathrm{nm}$) to $x_p = 15 \, \%$ ($r^{\mathrm{rem}}_{\mathrm{max}} \cong 15 \, \mathrm{nm}$). By contrast, the position of the first peak for the \textit{saturated} microstructures, $r^{\mathrm{sat}}_{\mathrm{max}} \cong 20.5 \, \mathrm{nm}$, is \textit{independent} of $x_p$. This difference in the $p(r)$-behavior between saturation and remancence gives rise to the shift in the oscillations of the corresponding $d \Sigma_M / d \Omega$ to larger $q$-values.

For the single Fe particle at $H_0 = 0$ we observe a minimum at around $r^{\mathrm{rem}}_{\mathrm{min}} \cong 29 \, \mathrm{nm}$ [Fig.~\ref{fig2}(b)]. While the position of this feature remains nearly constant with increasing concentration ($r^{\mathrm{rem}}_{\mathrm{min}} \cong 29-31 \, \mathrm{nm}$), the value of $p(r^{\mathrm{rem}}_{\mathrm{min}})$ evolves from negative to positive values when going from the single particle to $x_p = 15 \, \%$. This finding is due to the increased interparticle correlations. The $p(r)$ of the $x_p = 15 \, \%$ sample is positive for all $r$-values, and we see a second local maximum in $p(r)$ for $r$-values larger than $\sim 40 \, \mathrm{nm}$ [Fig.~\ref{fig3}(g)], which is related to the nearest-neighbor distance (first coordination shell); see also~\cite{laura2019sm} for the pair-distribution function. The shift of the second local maximum between saturation and remanence is attributed to the nonuniform internal spin structure.

{\it Conclusion.} Using micromagnetic computations we have investigated the transition from single-domain to multi-domain behavior in defect-free magnetic Fe nanoparticles and the related implications for the magnetic small-angle neutron scattering cross section $d \Sigma_M / d \Omega$ and pair-distance distribution function $p(r)$. We have demonstrated that the $d \Sigma_M / d \Omega$ and $p(r)$ of nonuniformly magnetized nanoparticles cannot be described anymore with the superspin model, which assumes a homogeneous spin microstructure. Away from saturation, deviations from the Guinier law and complicated real-space correlations are encountered. Increasing interparticle interactions modify the characteristics of $p(r)$, however, at the remanent state we have observed for all concentrations a reduction of the characteristic spin-structure size, which is related to the vortex-type inhomogeneous magnetization distribution of the Fe nanoparticles. This feature [shift of the main peak in $p(r)$] can be used by experimenters as an indication for the occurrence of an inhomogeneous magnetization texture (see also Fig.~7 in \cite{laura2019sm} for an experimental example).

The micromagnetic approach to magnetic SANS consist of finding, by means of magnetic-energy minimization, the three-dimensional vector field of the magnetization. This represents a paradigm shift and is conceptually very much different than the up-to-now used approach of finding a scalar function describing the magnetization profile of the particle ensemble based on well-behaving structural forms. Currently, no analytic description for the $d \Sigma_M / d \Omega$ and $p(r)$ of an inhomogeneously magnetized particle (sphere), or an interacting particle ensemble, is available, which represents a challenge for future studies. In analogy to the ongoing efforts in nuclear SANS and small-angle x-ray scattering on complex-shaped biological macromolecules~\cite{Franke:ge5042}, the compilation of a library of SANS cross sections and associated correlation functions for different nanoparticle sizes and shapes, size-distribution functions, packing densities of nanoparticles, different symmetries of magnetic anisotropy, defect structures, Dzyaloshinskii-Moriya interaction, etc. would be highly desirable. We believe that the combination of magnetic SANS with micromagnetic computations will allow one to reconstruct the three-dimensional spin structure of nanoparticles.

L.\ G.\ Vivas, P.\ Bender, and A.\ Michels ack\-now\-ledge financial support from the National Research Fund of Luxembourg (AFR Postdoc Grant No.~8865354 and CORE SANS4NCC grant). R.\ Yanes acknowledges financial support from Junta de Castilla y Leon (Project No.~SA090U16). This research was supported by the EU-H2020 AMPHIBIAN Project No.~720853.

\bibliographystyle{apsrev4-1}

\begin{thebibliography}{41}
\bibitem [{ser()}]{sergeycommentsd}%
  \BibitemOpen
  \href@noop {} {}\bibinfo {note} {We would like emphasize that this expression
  is valid for a single isolated spherical particle. The presence of dipolar
  interactions between the particles may lead to strong deviations from this
  simple estimate.}\BibitemShut {Stop}%
\end{thebibliography}

%

\end{document}